\documentclass[11pt,twoside]{article}

\newcommand{\M}{{\cal M}}
\newcommand{\D}{{\cal D}^{12}}
\newcommand{\Z}{{\cal Z}}
\newcommand{\C}{{\bf C}} 
\newcommand{\Pw}{{\cal P}}
\newcommand{\ra}{\rightarrow}
\newcommand{\nn}{\nonumber}

\leftline{\copyright~ 1999 International Press}
\leftline{Adv. Theor. Math. Phys. {\bf 2} (1998) {\ 1287 -- 1306} }

\begin{document}
%\begin{titlepage}
%\renewcommand{\thepage}{ }
%\renewcommand{\today}{ }
\thispagestyle{empty}
%\maketitle

\begin{center}
\vspace{0.2in}
{\huge \bf $D_k$ Gravitational Instantons \\[5mm] and Nahm Equations}
\vspace{0.2in}
\renewcommand{\thefootnote}{}
\footnotetext{\small e-print archive: {\texttt
http://xxx.lanl.gov/abs/hep-th//9803112}}
\renewcommand{\thefootnote}{*}
\footnotetext{Research supported in part by DOE grant DE-FG03-92-ER40701}
\renewcommand{\thefootnote}{**}
\footnotetext{Research supported in part by DOE grant DE-FG02-90ER40542}
\renewcommand{\thefootnote}{\arabic{footnote}}

{\bf Sergey A.  Cherkis$^{*}$,  Anton Kapustin$^{**}$}
\vspace{0.2in}

$^{*}$California Institute of Technology\\
Pasadena,\\ 
CA 91125
\vspace{0.1in}

$^{**}$School of Natural Sciences,\\
Institute for Advanced Study\\
Olden Lane, Princeton \\
NJ 08540
\vspace{0.1in}

\end{center}
\begin{abstract}
We construct $D_k$ Asymptotically Locally Flat gravitational instantons
as moduli spaces of solutions of Nahm equations. This allows us to find
their twistor spaces and K\"ahler potentials.
\end{abstract}

\section{Introduction} 
Gravitational instantons are four-dimensional manifolds with a self-dual
curvature tensor. They can be compact (K3 and $T^4$) or noncompact. 
Asymptotically Locally Euclidean (ALE) instantons are gravitational instantons that approach
${\bf R}^4/\Gamma$ at infinity, where $\Gamma$ is a finite subgroup of $SU(2)$.
These were constructed by Gibbons and Hawking~\cite{GH} for $\Gamma={\bf Z}_k$ and by Kronheimer 
\cite{ALE} for all possible $\Gamma$. 
\newpage
\pagenumbering{arabic}
\setcounter{page}{1288}

\pagestyle{myheadings}
\markboth{\it $D_k$ GRAVITATIONAL INSTANTONS AND NAHM EQUATIONS}{\it S. A. CHERKIS, A. KAPUSTIN}

The purpose of this paper is to present a construction of a class of Asymptotically Locally
Flat (ALF) spaces. At infinity the metric on these spaces approaches
\begin{equation}
ds^2=dr^2+\sigma_1^2+r^2(\sigma_2^2+\sigma_3^2) ,
\end{equation}
where $\sigma_j$ are left-invariant one-forms on ${\bf S}^3/\Gamma$ for some
finite subgroup $\Gamma$ of $SU(2)$. 
Thus these spaces are classified by a choice of $\Gamma$ which can be one of the following groups:
${\bf Z}_k$ (cyclic), ${\bf D}_k$ (binary dihedral), ${\bf T}$ (binary tetrahedral), 
${\bf O}$ (binary octahedral), ${\bf I}$ (binary icosahedral). The first two are infinite 
classes parametrized by an integer $k$. It is known~\cite{ALE} that in the ALE case all of 
these possibilities are realized. The corresponding spaces are called $A_{k-1}, D_k, E_6, 
E_7,$ and $E_8$ ALE spaces, respectively. The names come from the McKay correspondence 
between finite subgroups of $SU(2)$ and simple Lie algebras. These names also reflect the 
topology of the respective spaces, as the intersection matrix of compact two
cycles of such a space is given by the negative of the Cartan matrix of the 
corresponding Lie algebra.

An example of the $A_k$ ALF space is described by the multi-Taub-NUT metric with $k+1$ centers.
In this paper we present the construction of $D_k$ ALF spaces.
Namely, we construct their twistor spaces, find all real holomorphic sections of
the latter, and compute the K\"ahler potential.

Our motivation comes from string theory. Let us consider M theory compactified on the $D_k$ ALF space.
As explained by Sen~\cite{Sen}, in IIA string theory
this corresponds to an orientifold 6-plane with $k$ parallel D6-branes. 
We can probe this background with a D2-brane parallel to the 6-branes.
The theory on the D2-brane is described at low energies by an N=4 $SU(2)$ 
Yang-Mills with $k$ fundamental hypermultiplets. Coordinates on the Coulomb
branch are the vevs of the Higgs fields and the dual photon.
{}From the M-theory point of view, the D2-brane is a membrane of M-theory, Higgs
fields correspond to the position of the membrane in the three-space transverse
to the six-branes, and the dual photon corresponds to the position of the membrane on the 
circle of M-theory. This allows one to identify a point on the ALF space
at which the membrane is positioned with a vacuum of the gauge theory on the D2-brane. 
So the Coulomb branch of the $D=3, N=4$
$SU(2)$ gauge theory with $k$ fundamental hypermultiplets should be a $D_k$ ALF space.
One can also see this from the gauge theory analysis~\cite{GAUGE}.

Remarkably, there is another realization of these gauge theories in type IIB string 
theory~\cite{HW}, as shown in Figure~1.

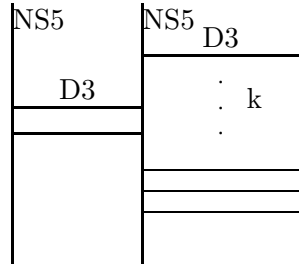
\begin{figure}[h!]
\setlength{\unitlength}{0.9em}
\begin{center}
\begin{picture}(17,13)
\put(3,2){\line(0,1){10}}
\put(8,2){\line(0,1){10}}
\multiput(3,7)(0,1){2}{\line(1,0){5}}
%\put(3,9){\line(1,0){5}}
%\multiput(5.5,7)(0,0.5){3}{\circle*{.1}}
\put(5.5,8.7){\makebox(0,0){D3}}
\put(8,11){NS5}
\put(3,11){NS5}

\multiput(8,4)(0,0.8){3}{\line(1,0){6}}
\put(8,10){\line(1,0){6}}
\multiput(11,7)(0,1){3}{\circle*{.1}}
\put(12,8){k}
\put(11,10.7){\makebox(0,0){D3}}

\end{picture}
\end{center}
\caption{NS5-branes are parallel to the $x^0, x^1, x^2, x^3, x^4, x^5$, and D3-branes are parallel
to the $x^0, x^1, x^2, x^6$ directions. $x^6$ is the horizontal direction on the figure.}
\end{figure}
In the infrared limit the theory on the internal D3-branes reduces to a 
$2+1$~dimensional theory. Namely, it is a $D=3,N=4$ $U(2)$ Yang-Mills theory
with $k$
fundamentals.
As discussed in Ref.~\cite{us1}, every vacuum of the gauge theory on the internal
D3-branes corresponds to a singular $SU(2)$ two-monopole on the NS5-branes with $k$ singularities.
These singularities look like pointlike Dirac monopoles embedded in $SU(2)$.
By virtue of Nahm transform, monopoles correspond to solutions of Nahm equations 
(the reduction of self-duality equations to one dimension).

Another way to see how Nahm equations appear is to look for supersymmetric vacua
of the $D=4,N=4$ theory on the D3-branes~\cite{Diaconescu}. 
The Higgs fields in a vacuum configuration depend on the $x^6$ coordinate 
(which we shall call $s$) and take values in the adjoint of $U(2)$ on the interval 
between the two NS5-branes and in the adjoint of $U(k)$ on the semiinfinite interval.
The equations that Higgs fields satisfy are nothing but Nahm 
equations~\cite{Diaconescu}.

Thus from string theory considerations it follows that the moduli space $\M$ of solutions
of Nahm equations corresponding to Figure~1 is a $D_k$ ALF gravitational instanton. 

The advantage of having a description in terms of Nahm equations is that
they can be interpreted as a moment map of a hyperk\"ahler quotient 
(see e.g., Ref.~\cite{Hitchin}). This guarantees that their
moduli space $\M$ is hyperk\"ahler. Furthermore, all three complex structures of $\M$
become apparent.

The paper is organized as follows:
In Section~2 we formulate the relevant Nahm data. In Section~3
we exhibit the moduli space $\M$ of Nahm data as a complex manifold with respect to one
of its complex structures.
In order to extract all the information about the hyperk\"ahler
manifold $\M$ (e.g., its metric) we need to find a whole 2-sphere of complex structures. 
In Section~4 we explore the dependence of the construction on the choice of the complex 
structure and find the twistor space $Z$ of $\M$. In order to find the metric on $\M$
one needs to find a decomposition $Z={\bf P}^1\times\M$. This is done in Section~5.
In Section~6 we obtain the K\"ahler potential for the $D_k$ ALF metric.
The final result agrees with a conjecture of Chalmers~\cite{Chalmers}.

\section{Nahm equations}
As explained in Refs.~\cite{Diaconescu,us1} the configuration of D3-branes in
Figure~1 can be described by certain matrix data called Nahm data.

The Nahm data will be a set of four functions $T_0, T_j,$ $j=1,2,3$ of real 
coordinate $s\in [-h,\infty)$ taking values in $u(2)$ for $s<0$ and in $u(k)$  
for $s>0$. The matching condition at $s=0$ is 
\begin{equation} \label{match}
T_j (s>0)=\left(\begin{array}{cc} -i\rho_j/2s+O(1) & 
                                        s^{(k-3)/2}p_j+O(s^{(k-1)/2})\\
                    -s^{(k-3)/2}\bar{p}_j^T+O(s^{(k-1)/2}) &
                                                T_j(0-)+O(s) 
                                                \end{array}\right) ,
\end{equation}
where $\rho_j$ are $k-2$ dimensional representations of Pauli sigma matrices
$\sigma_j$, $p_j$ are $(k-2)\times2$ matrices.
Near $s=-h$ we require $T_j(s)=-\frac{i}{2}\frac{\sigma_j}{s+h}+O(s+h)$,
and at $s\rightarrow\infty$ \ $T_j\rightarrow diag(x^{(j)}_1, x^{(j)}_2, \ldots, x^{(j)}_k)$. 
The parameter $h$ is the distance between the NS5-branes in Figure~1.
We will see that $h$ is inversely proportional to the radius of the circle at the
asymptotic infinity of the ALF space.

Nahm data are acted upon by a gauge group $G$. $G$
is defined as a group whose elements are functions $g(s)$ valued in $U(2)$ for $s\in [-h, 0]$ 
and in $U(k)$ for $s>0$,
and satisfying $g(-h)=1$, $g(+\infty)=1$, and
\begin{equation} \label{gauge}
g(0+)=\left(\begin{array}{cc} 1 & 0 \\ 0 & g(0-)\end{array}\right).
\end{equation} 

We subject the Nahm data to Nahm equations
\begin{equation} \label{Nahm}
\frac{d T_i}{ds}+[T_0,T_i]=\frac{1}{2}\epsilon_{ijk}[T_j,T_k].
\end{equation}
The moduli space of such Nahm data modulo gauge transformations is the 
hyperk\"ahler manifold $\M$ we are interested in.

If we consider all quadruplets $T_0, T_1, T_2, T_3$
valued in $u(2)$ on $[-h,0)$ and in $u(k)$ on $[0,\infty)$, we can define 
a norm  
\begin{equation}
||T||^2=Tr\int_{-h}^{\infty} \left(T_0^2+T_1^2+T_2^2+T_3^2\right) ds.
\end{equation}
The metric corresponding to this norm is hyperk\"ahler. This becomes obvious
if we think of a quadruplet $T_0,\ldots,T_3$ as a matrix of quaternions 
$$T=T_0+i T_1+j T_2+k T_3.$$ Then the Nahm equations Eq.~(\ref{Nahm}) are moment maps 
for the action of $G$
\begin{equation}
T_0\rightarrow g^{-1}T_0 g+g^{-1}dg/ds,\ T_j=g^{-1} T_j g.
\end{equation}

We construct the moduli space $\M$ in the following way: 
Let us first consider the restriction of the Nahm data
to the interval $[0,\infty)$ modulo the gauge transformations with $g(0)=1$ and 
call the resulting manifold $\M_{+}$. Then consider the restriction of the
Nahm data to $[-h,0]$ modulo the gauge transformations with $g(0)=g(-h)=1$ 
and call this manifold $\D$. We may relax the condition $g(-h)=1$ and consider 
the gauge transformations with $g(-h)\sim 1$.
This gives a triholomorphic action of $U(1)$ on $\D$. We may think of this $U(1)$
as the group of gauge transformations ``localized'' at $s=-h$. The group of all 
gauge transformations modulo those with $g(0)=1$ is $U(2)$. This $U(2)$ can be thought of 
as ``localized'' at $s=0$. It acts triholomorphically on both $\M_+$ and $\D$. 
The manifold $\M$ is a hyperk\"ahler quotient of $\M_+\times\D$ by $U(1)\times U(2)$
at zero level.

\section{The complex structure of $\M$}      \label{sec:complex}
Following Donaldson~\cite{Donaldson} we define
\begin{equation}\label{alphabeta}
\alpha=\frac{1}{2}\left(T_0-i T_1\right), \ \beta=\frac{1}{2}\left(T_2+i T_3\right).
\end{equation}
Then the Nahm equations can be written as a pair of equations. We will need only one of
them called the ``complex'' equation: 
\begin{equation}  \label{complex}
\frac{d\beta}{ds}+2[\alpha, \beta]=0.
\end{equation}
Denote by $\M$ the space of Nahm data satisfying Nahm equations 
modulo the group $G$ of gauge transformations. Let $\tilde{\M}$ be the space 
of solutions of the complex equation Eq.~(\ref{complex}) modulo the complexified
group of gauge transformations $G^{\C}$. One can show~\cite{Donaldson} that $\M$ as a comlpex variety is the same as $\tilde{\M}$. As we will see it is rather easy to solve
Eq.~(\ref{complex}). This allows us to describe $\M$ as a complex manifold.

%{\bf New boundary conditions}

First let us look at $\M_{+}$.
Let $a=s\alpha, b=s\beta, s=e^t$; then the complex equation Eq.~(\ref{complex})
is equivalent to
\begin{equation} \label{ab}
\frac{db}{dt}+2[a,b]=b.
\end{equation}
Fixing the gauge at
$t\rightarrow-\infty$ so that $a$ is independent of $t$ we can solve Eq.(\ref{ab})
as follows:
\begin{equation}
a=\left(\begin{array}{cc} -\rho_1/4 & 0 \\ 0 & 0 \end{array}\right),\qquad
b=e^t \left(\begin{array}{cc} e^{\rho_1 t/2} & 0 \\ 0 & 1 \end{array}\right)
\hat{b} \left(\begin{array}{cc} e^{-\rho_1 t/2} & 0 \\ 0 & 1 
                                                        \end{array}\right).
\end{equation}
Here the matrix $\hat{b}$ is independent of $t$. Comparing with the boundary condition 
at $t\rightarrow-\infty$ following from Eq.~(\ref{match}), we find
\begin{equation}
\alpha=\left(\begin{array}{cc} -\frac{1}{4 s}\rho_1 & 0 \\ 
                                                0 & 0 \end{array}\right),\qquad
\beta=\left(\begin{array}{cc} -\frac{i}{4s}\left(\rho_2+i \rho_3\right)+A &
                                        (-2 i s)^{(k-3)/2} \tilde{p}\\
(-2 i s)^{(k-3)/2} \tilde{q} & B \end{array} \right) ,
\end{equation}
where the constant matrices $A,\tilde{p},\tilde{q}$ satisfy 
$[\rho_1,A]=0,\ \rho_1 \tilde{p}=(k-3) \tilde{p},\ \tilde{q}\rho_1=-(k-3)\tilde{q}$. 
It follows that $A$ is a diagonal $(k-2)\times (k-2)$ matrix (in the basis in which $\rho_1$
is diagonal). We call its eigenvalues
$A_1,\ldots,A_{k-2}$. Now recall that $\tilde{p}$ and $\tilde{q}$
are $(k-2)\times2$ and $2\times(k-2)$ matrices. Letting a vector $v$ be the highest weight of the
representation $\rho$ (that is $\rho_1 v=(k-3) v$) with $|v|=1$, we see that
the two columns of $\tilde{p}$ are $p_1 v$ and $p_2 v$ where $p_1,p_2\in {\bf C}$.
So $\tilde{p}$ is parametrized by a vector $p\in {\bf C}^2$ with components $(p_1, p_2)$.
Similarly, $\tilde{q}$ can be parametrized by a vector $q\in {\bf C}^2$.

Boundary conditions at $s\rightarrow+\infty$ restrict the eigenvalues of $\beta$
to be $z_a=(x^{(2)}_a+i x^{(3)}_a)/2$. Thus the equation $\det(\beta-z)=0$ has roots $z_a$.
On the other hand, a direct computation of the determinant yields
\begin{equation}\label{eigcond}
\det(\beta-z)=\det (B-z) \left(\det(A-z)-p^T (B-z)^{-1} q\right).
\end{equation}

%{\bf Better description of Dancer Manifold}

Now let us turn our attention to $\D$. The solutions of Nahm equations on $s\in[-h,0]$ 
modulo gauge transformations with precisely the right boundary conditions were described by 
Dancer~\cite{Dancer0,Dancer}. As a complex manifold, $\D$ is a set of pairs $(B,w)$ with $B$ being a 
$2\times2$ matrix and $w\in {\bf C}^2$, such that $w$ and $Bw$ are linearly independent.

%{\bf centering => Tr B=0}

So far we described the solutions of the complex equation Eq.~(\ref{complex}) modulo 
complexified gauge transformations $g(s)$ satisfying $g(0)=g(-h)=1$. 
Every such solution is given by a set $(A,B,p,q,w)$ where $A$ is a diagonal
matrix with eigenvalues $A_1,\ldots,A_{k-2}$, $B$ is a $2\times 2$ matrix, and 
$p,q,w\in {\bf C}^2$. Now we have to take a symplectic quotient by the complexified groups
$U(1)^{\C}=\C^*$ and $U(2)^{\C}\simeq\C^*\times SL(2,\C)$. The moment map for the action
of $U(1)^{\C}$ is $Tr\ B$, so we require $Tr\ B=0$.

An element $g$ of $SL(2,\C)$ acts on these data as
\begin{equation} \label{glue}
B\rightarrow g B g^{-1},\quad w\rightarrow g w,\quad p^T\rightarrow p^T g^{-1},\quad 
q\rightarrow g q,
\end{equation}
and the two $\C^*$ actions (we will call them $\C^*_{\lambda}$ and $\C^*_{\kappa}$)
are given by
\begin{eqnarray}
&B\rightarrow B,\quad w\rightarrow  w,\quad p^T\rightarrow p^T \lambda^{-1}, \quad
q\rightarrow \lambda q&,\\
&B\rightarrow B,\quad w\rightarrow \kappa w,\quad p^T\rightarrow p^T,\quad 
q\rightarrow q.&
\end{eqnarray}
The $\C^*_{\kappa}$ action and the fact that $w$ and $Bw$ are linearly independent allows us
to put $$w^T \left(\begin{array}{cc} 0&1\\-1&0\end{array}\right)B w=1.$$ This gauge-fixes
$\C^*_{\kappa}$. Now we can define $SL(2,\C)$ invariants
$x_1, x_2,$ $y_1,$ and $y_2$ by 
\begin{equation}
p^T\left(\begin{array}{cc} 0&1\\-1&0\end{array}\right)=x_1w^T B^T+x_2 w^T ,
\end{equation}
\begin{equation}
\ q =-y_1 B w+y_2 w.
\end{equation} 
$x_1,x_2,y_1,y_2,\det B$ and $A_1,\ldots,A_{k-2}$ form the full set of invariants
with respect to $SL(2,\C)$.

The residual $\C^*_\lambda$ acts by
$B\ra B, x_i\ra\lambda^{-1} x_i, y_i\ra\lambda y_i$.
Thus the invariants of $\C^*_\lambda\times SL(2,\C)$ are given by
\begin{eqnarray}
\eta_2&=&-\det B\nn\\
\Psi_1&=&x_2y_2\nn\\
\Psi_2&=&x_1y_1\nn\\
\Psi_3&=&x_1y_2\nn\\
\Psi_4&=&x_2y_1,
\end{eqnarray}
and $A_b$, $b=1,\ldots, k-2$. These invariants satisfy additional relations 
\begin{equation}\label{eq1}
\Psi_1\Psi_2=\Psi_3\Psi_4
\end{equation}
\begin{equation}\label{eq2}
\left(z_a^2-\eta_2\right)\prod_{b=1}^{k-2}\left(A_b-z_a\right)+
z\left( \Psi_3-\Psi_4\right)+\Psi_1-\Psi_2\eta_2=0
\end{equation}
for $a=1,\ldots,k$. The first of these follows from the definition of the invariants,
and the second one is a consequence of Eq.~(\ref{eigcond}).

Let us denote symmetric polynomials of order $m$ by $S_m$. For example
$S_1(z)=z_1+z_2+\ldots+z_k$, and $S_{k-2}(A)=A_1 A_2\ldots A_{k-2}$.
One can rewrite Eqs.~(\ref{eq1},\ref{eq2}) as
\begin{eqnarray*} \label{system}\
S_1(z)&=&S_1(A) \\
S_2(z)&=&S_2(A)-\eta_2 \\
S_3(z)&=&S_3(A)-\eta_2 S_1(A) \\
\vdots&&\\
S_m(z)&=&S_m(A)-\eta_2 S_{m-2}(A) \\
\vdots&&\\
S_{k-2}(z)&=&S_{k-2}(A)-\eta_2 S_{k-4}(A)\\
\Psi_1&=&\eta_2 \Psi_2+F(\eta_2)\\
\Psi_4&=&\Psi_3+G(\eta_2)
\end{eqnarray*}
\begin{equation} \label{dk}
\eta_2 \Psi_2^2+\Psi_2 F(\eta_2)=\Psi_3^2+\Psi_3 G(\eta_2) ,
\end{equation}
where the polynomials $F(\eta_2)$ and $G(\eta_2)$ are defined as
\begin{eqnarray}\label{dk1} 
F(\eta_2)&=&S_k(z)+S_{k-2}(z)\eta_2+\ldots+S_{k-2l}(z)\eta_2^l+\ldots \nn\\
G(\eta_2)&=&S_{k-1}(z)+S_{k-3}(z)\eta_2+\ldots+S_{k-1-2l}(z)\eta_2^l+\ldots.
\end{eqnarray}
The above equations define our complex manifold $\M$ as a subvariety
in $\C^{k+3}$.
{}From Eqs.~(\ref{dk},\ref{dk1}) one can see that $\M$ develops a 
$D_k$-type singularity when all $z_a$ are set to zero.

\section{The Twistor Space of $\M$}
In the previous section we described $\M$ as a complex variety by
picking a particular complex structure on the space of Nahm data. In reality there
is a whole 2-sphere of such complex structures. To obtain the twistor space of
$\M$ we need to trace the dependence on the choice of the complex structure.
 
Let $\zeta$ be an affine parameter on the ${\bf CP}^1$ of complex structures. We define
\begin{eqnarray}
\alpha^0&=&\frac{1}{2}\left(T_0+i T_1+\zeta\left(T_3-i T_2\right)\right),\nonumber \\ 
\beta^0&=&\frac{i}{2}\left(T_3+i T_2+2 i \zeta T_1+
\zeta^2\left(T_3-i T_2\right)\right)
\end{eqnarray}
for $\zeta\neq\infty $ and
\begin{eqnarray}
\alpha^1&=&\frac{1}{2}\left(T_0-i T_1-\frac{1}{\zeta}\left(T_3+i T_2\right)\right), \nonumber \\ 
\beta^1&=&\frac{1}{2}\left(T_2+i T_3-2 \frac{1}{\zeta} T_1+
\frac{1}{\zeta^2}\left(-T_2+i T_3\right)\right)
\end{eqnarray}
for $\zeta\neq 0$. Both pairs $(\alpha, \beta)$ satisfy the complex Nahm equation
\begin{equation}
\frac{d \beta}{ds}+2[\alpha,\beta]=0.
\end{equation}
The relation between them is given by
\begin{equation}          \label{tuda-suda}
\alpha^1=\alpha^0+\frac{i}{\zeta}\beta^0,\ \ \beta^1=\frac{1}{\zeta^2}\beta^0.
\end{equation}

Following the same steps as in the Section~\ref{sec:complex} and fixing the 
gauge so that $s \alpha^0,s\alpha^1$ are constant, we obtain
\begin{eqnarray}\label{alpha0}
\alpha^0&=&\left(\begin{array}{cc} 
        \frac{1}{4s} \left(\rho_1-2\zeta\rho_{+}\right)& 0\\
        0&0\end{array}\right),\\
\beta^0&=&\left(\begin{array}{cc}
   \frac{i}{2 s}\left(\rho_{-}+\zeta\rho_1-\zeta^2\rho_{+}\right)+
   e^{-\zeta\rho_{+}}A^0(\zeta) e^{\zeta\rho_{+}}& s^{(k-3)/2} e^{-\zeta\rho_{+}} p^0(\zeta)\\
   s^{(k-3)/2} q^0(\zeta) & B^0(\zeta) \end{array}\right)\nn
\end{eqnarray}
and
\begin{eqnarray}\label{alpha1}
\alpha^1&=&\left(\begin{array}{cc} 
        -\frac{1}{4s} \left(\rho_1+\frac{2}{\zeta}\rho_{-}\right)& 0\\
        0&0\end{array}\right),\\
\beta^1&=&\left(\begin{array}{cc}
   -\frac{i}{2 s}\left(\rho_{+}-\frac{1}{\zeta}\rho_1-\frac{1}{\zeta^2}\rho_{-}\right)+
   e^{-\rho_{-}/\zeta}A^1(\zeta) e^{\rho_{-}/\zeta}& s^{(k-3)/2} p^1(\zeta)\\
   s^{\frac{k-3}{2}} q^1(\zeta)e^{\frac{\rho_{-}}{\zeta}} & B^1(\zeta) \end{array}\right).\nn
\end{eqnarray}
In these formulas $\rho_{+}=\frac{1}{2} (\rho_2+i\rho_3), 
\rho_{-}=\frac{1}{2} (\rho_2-i\rho_3)$, 
\begin{eqnarray}
&[A^0(\zeta), \rho_1]=0,\ [A^1(\zeta),\rho_1]=0&\\
&\rho_1 p_0(\zeta)=-(k-3)p_0(\zeta),\ q_0(\zeta)\rho_1=(k-3) q_0(\zeta)&\\
&\rho_1 p_1(\zeta)=(k-3)p_0(\zeta),\ q_1(\zeta)\rho_1=-(k-3) q_1(\zeta).&
\end{eqnarray}

Equation (\ref{tuda-suda}) relates $(\alpha^0, \beta^0)$ and 
$(\alpha^1, \beta^1)$, but in a gauge different from that in Eqs.~(\ref{alpha0},\ref{alpha1}).
Namely the gauge 
transformation that makes $s\alpha^1=s\left(\alpha^0+\frac{i}{\zeta}\beta^0\right)$ 
independent of $s$ is
\begin{equation}
g=\exp\left(-\frac{2i}{s \zeta}\beta^0\right)
\exp\left(-\frac{1}{\zeta}\rho_{-}-\rho_1+\zeta \rho_{+}\right).
\end{equation}
Comparing the two expressions for $\beta^1$ we can find the transition functions
between $(A^0, p^0, q^0, B^0)$ and $(A^1, p^1, q^1, B^1)$. If we let
\begin{equation}
R=e^{\rho_{-}/\zeta}e^{-\zeta\rho_{+}}e^{\rho_{-}/\zeta}\nn
\end{equation}
then we get 
\begin{equation}
A^1=R A^0 R^{-1},\quad p^1=\frac{1}{\zeta^2} R p^0,\quad q_1=\frac{1}{\zeta^2}q^0 R^{-1},
\quad B^1=\frac{1}{\zeta^2} B^0.
\nn\end{equation}
This means that the variables $A,B,p,q$ are sections of the following bundles:
\begin{eqnarray}
B &{\rm of}& Mat(2)\otimes{\cal O}(2),\nn\\
A_b &{\rm of}& {\cal O}(2),\nn\\
p &{\rm of}& {\cal O}(k-1)\times {\cal O}(k-1),\nn\\
q &{\rm of}& {\cal O}(k-1)\times {\cal O}(k-1).
\end{eqnarray}
Here ${\cal O}(n)$ is defined as the line bundle on ${\bf P}^1$ with the transition 
function $\zeta^{-n}$. Furthermore, from Dancer's analysis~\cite{Dancer} it follows 
that $w^1=\zeta e^{2i h B^0/\zeta} w^0$.
{}From the above one can find the dependence of the invariants $\Psi_1,\ldots, \Psi_4$ on
$\zeta$. This completely determines the twistor space $\Z$ of $\M$. 

It turns out that the invariants $\Psi_i$ are not taking values in any nice fibrations,
but one can define certain combinations of them that do. Let $\eta=\sqrt{-\eta_2}$, 
and define 
\begin{eqnarray} \label{greek}
\mu&=&\Psi_1+\eta^2\Psi_2-i\eta(\Psi_3-\Psi_4),\nn\\
\nu&=&\Psi_1+\eta^2\Psi_2+i\eta(\Psi_3-\Psi_4),\nn\\
\rho&=&\Psi_1-\eta^2\Psi_2+i\eta(\Psi_3+\Psi_4),\nn\\
\xi&=&\Psi_1-\eta^2\Psi_2-i\eta(\Psi_3+\Psi_4).
\end{eqnarray}
These are two-valued functions of $\zeta$, because $\eta(\zeta)=\sqrt{-\eta_2(\zeta)}$,
but they have simple transformation properties:
\begin{eqnarray}
\tilde{\mu}&=&\zeta^{-2k}\mu,\ \tilde{\nu}=\zeta^{-2k}\nu,\\
\tilde{\rho}&=&\zeta^{-2k} e^{2h\eta/\zeta}\rho,\label{rhotrans} \\
\tilde{\xi}&=&\zeta^{-2k} e^{-2h\eta/\zeta}\xi.\label{xitrans}
\end{eqnarray}
Furthemore, from Eqs. (\ref{system}) we get
\begin{eqnarray}
\mu&=&\prod_{a=1}^k\left(z_a(\zeta)+i\eta\right),\\
\nu&=&\prod_{a=1}^k\left(z_a(\zeta)-i\eta\right),
\end{eqnarray}
and
\begin{equation} \label{main}
\rho \ \xi=\prod_{a=1}^k\left(z_a^2(\zeta)+\eta^2\right).
\end{equation}

From this description of $\Z$ one can see that $\Z$ is exactly 
the twistor space of the centered moduli space of two monopoles with $k$ singularities computed
in Ref.~\cite{us2}.
In the direct image sheaf construction of Ref.~\cite{us2}
$\rho_0\xi_0, \rho_1\xi_1, \rho_0\xi_1,$ and $\rho_1\xi_0$ correspond to
$\Psi_1, \Psi_2, \Psi_3, $ and $\Psi_4$. This establishes the isometry between the moduli
space $\M$ of Nahm data and the centered moduli space of singular monopoles of nonabelian 
charge two. As explained in the Introduction, the equivalence of the two
moduli spaces follows from string theory.

The twistor space $\Z$ is nothing but ${\bf P}^1\times\M$, so the fiber of $\Z$ over $\zeta$
is our manifold $\M$ with the complex structure determined by $\zeta$. In 
order to find the Riemannian metric of $\M$ we need to pick two local complex
coordinates on $\M$ that depend on $\zeta$ holomorphically. For example, locally
we can pick $(\eta, \rho)$ or $(\eta, \xi)$ as such coordinates. There is a 
natural two-form $\omega=2d\eta\wedge d\xi/\xi=-2d\eta\wedge d\rho/\rho$ on $\Z$.
This two-form is necessary to recover the metric on $\M$ from the twistor space 
$\Z$~\cite{AH}.
It is degenerate along the $\zeta$ direction and 
satisfies $\tilde{\omega}=\zeta^{-2}\omega$. Let us rewrite it as
\begin{equation}
\omega=d\eta\wedge d \log\frac{\xi}{\rho}.
\end{equation}
We can introduce a new coordinate 
\begin{equation} \label{chi}
\chi=\frac{1}{\eta} \log \frac{\rho}{\xi}=\frac{1}{\eta} \log\frac{\Psi_1-\eta^2\Psi_2-i\eta(\Psi_3+\Psi_4)}
{\Psi_1-\eta^2\Psi_2+i\eta(\Psi_3+\Psi_4)} ,
\end{equation} 
which does not depend on the choice of the branch of the square root in $\eta=\sqrt{-\eta_2}$. In terms
of $\eta_2$ and $\chi$, $\omega=d\eta_2\wedge d\chi.$ To compute the metric from the twistor
data we will employ the generalized Legendre transform of Ref.~\cite{LR,IR}. This technique
yields directly the K\"ahler potential of the metric. But first we have
to find $\chi$ as a function on $\zeta$. This is the subject of the next section.

\section{Real sections of $\Z$}
Let us recall that
$\eta_2=-\det B$ and $\tilde{B}=\zeta^{-2} B$, so $\eta_2$ is a 
quartic polynomial in $\zeta$. Also $\eta_2$ should satisfy $\eta_2(-1/\bar{\zeta})=
\bar{\zeta}^{-4}{\bar{\eta}}_2(\zeta)$. Thus 
\begin{equation}
\eta_2=z+v\zeta+w\zeta^2-\bar{v}\zeta^3+\bar{z}\zeta^4 ,
\end{equation}
where $z,v\in {\bf C}, w\in {\bf R}$.
Consider a curve $S$ given by $\eta^2+\eta_2(\zeta)=0$. It covers the ${\bf P}^1$ parametrized by $\zeta$ 
twice, and has genus one. We will think of it as a torus $\C/{\bf Z}^2$ with a holomorphic
coordinate $u$. 

To establish the connection between $\eta, \zeta$ and the coordinate $u$ let us make a change 
of variables
\begin{equation}
\zeta=\frac{a\hat\zeta+b}{-\overline{b}\hat\zeta+\overline{a}},\qquad
\eta=\frac{\hat\eta}{(-\overline{b}\hat\zeta+\overline{a})^2},\qquad |a|^2+|b|^2=1 ,
\end{equation}
so that the equation for $S$ takes the canonical form 
\begin{equation}\label{st}
\hat\eta^2=4k_1^2\left(\hat\zeta^3-3k_2\hat\zeta^2-\hat\zeta\right), k_1>0,k_2\in{\bf R}.
\end{equation}
{}From this form we see that one period of the torus, $\omega_r$, is real, and
the other one, $\omega_i$, is purely imaginary. So $S=\C/\Omega$ where $\Omega$ is a rectangular
lattice spanned by $\omega_r, \omega_i$. Every such curve $S$ is parametrized by an $SU(2)$ matrix 
$$\left(\begin{array}{cc} a&b\\-\bar{b}&\bar{a} \end{array}\right)$$
and two real numbers $k_1$ and $k_2$. The relation of $\hat{\eta}$ and $\hat{\zeta}$ with
$u$ is given by $\hat{\eta}=k_1\Pw^{'}(u), \hat{\zeta}=\Pw(u)+k_2$, where $\Pw(u)$ is the
Weierstrass elliptic function.

Now we regard $\rho$ and $\xi$ as functions of $u$. 
The equation $$\rho\ \xi=\prod_{\alpha=1}^k \left(z_a^2(\zeta)+\eta^2\right)$$ and the fact 
that $\rho$ and $\xi$ are interchanged by the change of the sign of $\eta$ implies that $\rho$ 
vanishes at the points $u_a$, $u_a^{'}$ in $u$ plane at which $\eta=z_a$, and $\xi$ vanishes
at the points $v_a$, $v_a^{'}$ at which $\eta=-z_a$.

Let us denote the two preimages of $\zeta=0$ in the $u$-plane by $u_0$ and $u_0^{'}$,
and the two preimages of $\zeta=\infty$ by $u_{\infty}$ and $u'_{\infty}$. Then $\rho/\xi$ is
meromorphic for $u\neq u_{\infty}, u_{\infty}^{'}$ with zeros at $u_a$, $u_a^{'}$
and poles at $v_a$, $v_a^{'}$, while 
$$\frac{\tilde{\rho}}{\tilde{\xi}}=e^{4h\eta/\zeta}\frac{\rho}{\xi}$$ 
is meromorphic for $u\neq u_0, u_0^{'}$.

Let us cover $S$ with two charts $u\neq u_{\infty}, u_{\infty^{'}}$ and
$u\neq u_0, u'_0$. In the first chart we find
\begin{equation}
\frac{\rho}{\xi}=\exp\left(-4h k_1\left(\zeta_w(u+u_{\infty})+\zeta_w(u-u_{\infty})\right)-2 c u\right)
\prod_a \frac{\sigma(u-u_a)\sigma(u-u'_a)}{\sigma(u-v_a)\sigma(u-v'_a)}.
\end{equation}
Here $\zeta_W(u)$ and $\sigma(u)$ are Weierstrass qua\-sielliptic functions 
(see e.g. Ref.~\cite{EllFunc} for definitions), 
and $c$ is a constant to be determined. The monodromy properties 
of the Weierstrass functions $\zeta_W$ and $\sigma$ are
\begin{equation}
\sigma(u+\omega_r)=-\sigma(u) e^{2\eta_r\left( u+\omega_r/2\right)} ,\
\sigma(u+\omega_i)=-\sigma(u) e^{2\eta_i\left( u+\omega_i/2\right)} , \nn
\end{equation}
\begin{equation}
\zeta_W(u+m\omega_r+n\omega_i)=\zeta_W(u)+2m\eta_r+2n\eta_i.\nn
\end{equation}
where $\eta_r=\zeta_W(\omega_r/2)$ and $\eta_i=\zeta_W(\omega_i/2)$.
Using the above relations one can find the transformation law of $\log\frac{\rho}{\xi}$ when $u$
changes by a period of the lattice $\Omega$.
In fact,  $\rho$ and $\xi$ have to be well-defined functions on the torus $S$, therefore the 
change in their logarithms should be an integer multiple of $2\pi i$ on every period of 
$\Omega$. This imposes a constraint on $S$ and allows us to determine the constant $c$. 
Considering $\log\rho$ yields
\begin{eqnarray}
\eta_r \left(-2 h k_1-\sum_a(u_a+u'_a)\right)-c_{\rho} \omega_r&=&\pi i n_r,\\
\eta_i \left(-2 h k_1-\sum_a(u_a+u'_a)\right)-c_{\rho} \omega_i&=&\pi i n_i,
\end{eqnarray}
where $n_i$ and $n_r$ are integers, and $c_\rho$ is a constant to be determined.
Using Legendre's relation $\eta_r\omega_i-\eta_i\omega_r=\pi i$ we have
\begin{equation}\label{dku}
-2 h k_1-\sum_a(u_a+u'_a)=n_r\omega_i-n_i\omega_r.
\end{equation}
Using the same reasoning for $\log\xi$, we obtain
\begin{equation}\label{dkv}
-2 h k_1+\sum_a(v_a+v'_a)=m_r\omega_i-m_i\omega_r.
\end{equation}
In order to find the integers $n_r, n_i, m_r,$ and $m_i$ we consider the limit
of large $k_1$ which corresponds to asymptotic infinity on our space $\M$.
In this limit the roots of the equation $\eta_2=z_a^2$ tend to the roots of $\eta_2=0$,
consequently $Re(u_a+u'_a)\rightarrow \omega_r$, $Re(v_a+v'_a)\rightarrow \omega_r$. 
The condition that Nahm data be nonsingular
inside the interval $s\in(-h,0)$ requires that there are no solutions to the
above equations for any smaller parameter $h$. This yields
\begin{equation}
m_i=k-1,\qquad n_i=-k-1,\qquad m_r=-k, \qquad n_r=k.
\end{equation}
Then the transformation properties of $\log\frac{\rho}{\xi}$ become
\begin{equation} \label{trans}
\log\frac{\rho(u+\omega_r)}{\xi(u+\omega_r)}=\log\frac{\rho(u)}{\xi(u)},
\qquad \log\frac{\rho(u+\omega_i)}{\xi(u+\omega_i)}=\log\frac{\rho(u)}{\xi(u)}+4\pi i.
\end{equation}
Adding Eqs.~(\ref{dku}) and (\ref{dkv}) together we obtain a constraint that the curve $S$ has
to satisfy:
\begin{equation} \label{constrain}
-4 h k_1+\sum_{a=1}^k(v_a+v'_a-u_a-u'_a)=2\omega_r.
\end{equation}
One can also determine $c$, but we do not need it here.

Recall that the curve $S$ was parametrized 
by two complex numbers $a,b$ satisfying $|a|^2+|b|^2=1$, and two real numbers $k_1, k_2$,
which makes a total of five parameters. Eq. (\ref{constrain}) reduces the number of
independent parameters to four. Thus we obtain a four-parameter family of real holomorphic
sections of $\Z$. The space of parameters is nothing but $\M$. In the next section we 
compute the K\"ahler potential for the metric on $\M$.

\section{K\"ahler potential} 
Having described the twistor space $\Z$ of $\M$ and its holomorphic sections
we now find the K\"ahler potential of $\M$. As mentioned before,
$\eta_2(\zeta)$ and $\chi(\zeta)$ (see Eq.~(\ref{chi}))
can be thought of as local complex coordinates on $\M$. Furthermore,
$\eta_2=z+v\zeta+w\zeta^2-\bar{v}\zeta^3+\bar{z}\zeta^4$ depends on $\zeta$
holomorphically, so we can apply the generalized Legendre transform construction of 
Ref.~\cite{LR,IR}. 
We briefly outline the construction here. We must compare the coordinate
$\chi$ holomorphic in the neighborhood of $\zeta=0$ with the coordinate $\tilde{\chi}$
holomorphic in the neighborhood of $\zeta=\infty$. To this end we define a function 
$\hat{f}$ and a contour $C$ by the equation
\begin{equation}\label{hat}
\oint_C \frac{d\zeta}{\zeta^n} \hat{f}=\oint_0 \frac{d\zeta}{\zeta^{n-2}}\chi-
\oint_{\infty} \frac{d\zeta}{\zeta^n} \tilde{\chi} ,
\end{equation}
which has to hold for all integer $n$.
Further, we define a function $G(\eta_2,\zeta)$ by
$\partial G/\partial \eta_2=\hat{f}/\zeta^2$ and consider a contour integral
\begin{equation}
F=\frac{1}{2\pi i}\oint_C \frac{d\zeta}{\zeta^2} G.
\end{equation}
$F$ is effectively a function of the coefficients $z,v$ and $w$ of the polynomial $\eta_2$.
According to Ref.~\cite{IR}, to obtain the K\"ahler potential $K$ one has to impose a 
constraint $\partial F/\partial w=0$ and 
perform Legendre transform on $F$ with respect to $v$ and $\bar{v}$:
\begin{equation}
K(z,\overline{z},t,\overline{t})=F(z,\overline{z},v,\overline{v},w)-tv-
\overline{t}\overline{v},\ \
\frac{\partial F}{\partial v}=t,\frac{\partial F}{\partial \overline{v}}=\overline{t}.
\end{equation}
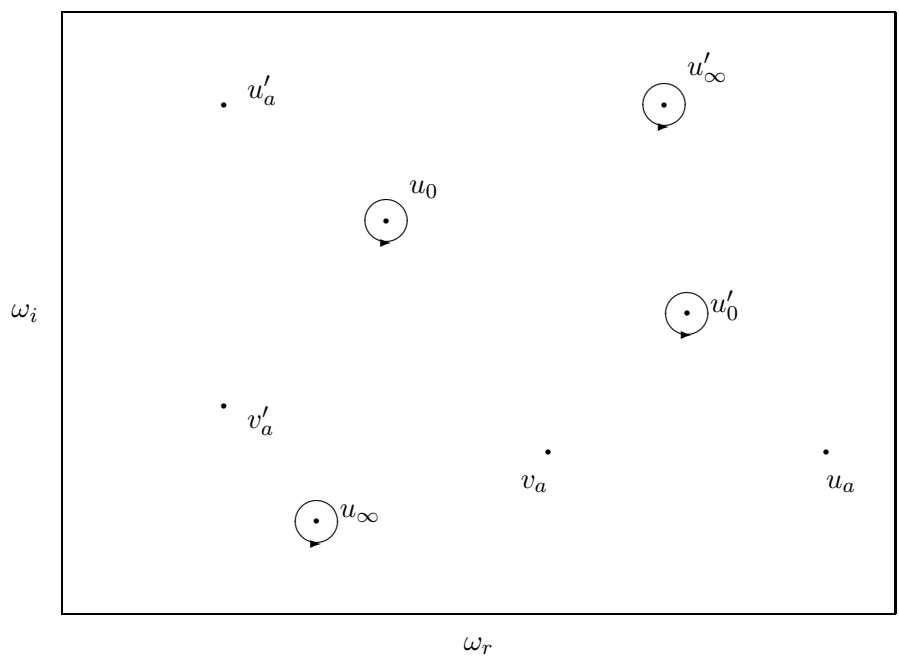
\begin{figure}
\setlength{\unitlength}{0.8em}
\begin{center}
\begin{picture}(40,30)
\put(2,2){\line(1,0){36}}
\put(38,2){\line(0,1){26}}
\put(38,28){\line(-1,0){36}}
\put(2,28){\line(0,-1){26}}

\put(9,11){\circle*{.2}}
\put(10,11){\makebox(0,0)[tl]{$v'_a$}}
\put(9,24){\circle*{.2}}
\put(10,24){\makebox(0,0)[bl]{$u'_a$}}
\put(13,6){\circle*{.2}}
\put(14,6){\makebox(0,0)[bl]{$u_{\infty}$}}
\put(16,19){\circle*{.2}}
\put(17,20){\makebox(0,0)[bl]{$u_0$}}
\put(23,9){\circle*{.2}}
\put(23,8){\makebox(0,0)[tr]{$v_a$}}
\put(35,9){\circle*{.2}}
\put(35,8){\makebox(0,0)[tl]{$u_a$}}
\put(29,15){\circle*{.2}}
\put(30,16){\makebox(0,0)[tl]{$u'_0$}}
\put(28,24){\circle*{.2}}
\put(29,25){\makebox(0,0)[bl]{$u'_{\infty}$}}

\put(13,6){\circle{2}} \put(13.2,5.05){\vector(1,0){0}}
\put(29,15){\circle{2}} \put(29.2,14.05){\vector(1,0){0}}
\put(16,19){\circle{2}} \put(16.2,18.05){\vector(1,0){0}}
\put(28,24){\circle{2}} \put(28.2,23.05){\vector(1,0){0}}

\put(20,1){\makebox(0,0)[t]{$\omega_r$}}
\put(1,15){\makebox(0,0)[r]{$\omega_i$}}

\end{picture}
\end{center}
\caption{Contour $\Gamma_1$.}
\end{figure}

Since we know the function $\chi$ explicitly as a function of $u$, it is more convenient to 
work with contour integrals in the $u$-plane, rather than in the $\zeta$-plane.
Let as rewrite Eq.(\ref{hat}) as
\begin{equation}\label{zerostep}
\oint_C \frac{d\zeta}{\zeta^n} \hat{f}=\left(\oint_0 \frac{d\zeta}{\zeta^{n-2}}\chi-
\oint_{\infty} \frac{d\zeta}{\zeta^{n-2}} \chi\right)+\left(\oint_
{\infty} \frac{d\zeta}{\zeta^{n}}\left(\zeta^2\chi-\tilde{\chi}\right)\right).
\end{equation}
The first step is to compute
\begin{equation}
 \oint_0 \frac{d\zeta}{\zeta^{n-2}}\chi-\oint_{\infty} \frac{d\zeta}{\zeta^{n-2}} \chi=
\frac{1}{2 k_1} \oint_{\Gamma_1}\frac{du}{\zeta^{n-2}(u)} \log\frac{\rho(u)}{\xi(u)} ,
\end{equation}
where the contour $\Gamma_1$ is shown in Figure~2.

We can deform $\Gamma_1$ to the contour $\Gamma_2$ in Figure~3.
\begin{figure}
\setlength{\unitlength}{0.8em}
\begin{center}
\begin{picture}(40,30)
%\put(8.5,0.5){\makebox(0,0){Figure~4}}
\put(2,2){\line(1,0){36}}
\put(38,2){\line(0,1){26}}
\put(38,28){\line(-1,0){36}}
\put(2,28){\line(0,-1){26}}

\put(9,11){\circle*{.2}}
\put(10,11){\makebox(0,0)[tl]{$v'_a$}}
\put(9,24){\circle*{.2}}
\put(10,24){\makebox(0,0)[bl]{$u'_a$}}
\put(13,6){\circle*{.2}}
\put(14,6){\makebox(0,0)[bl]{$u_{\infty}$}}
\put(16,19){\circle*{.2}}
\put(17,20){\makebox(0,0)[bl]{$u_0$}}
\put(23,9){\circle*{.2}}
\put(23,8){\makebox(0,0)[tr]{$v_a$}}
\put(35,9){\circle*{.2}}
\put(35,8){\makebox(0,0)[tl]{$u_a$}}
\put(29,15){\circle*{.2}}
\put(30,16){\makebox(0,0)[tl]{$u'_0$}}
\put(28,24){\circle*{.2}}
\put(29,25){\makebox(0,0)[bl]{$u'_{\infty}$}}

\put(9,17.5){\oval(2,15)} \put(8,18){\vector(0,1){0}}
\put(29,9){\oval(14,2)} \put(29,10){\vector(1,0){0}}

\put(20,2){\vector(1,0){0}}\put(38,15){\vector(0,1){0}}
\put(20,28){\vector(-1,0){0}}\put(2,15){\vector(0,-1){0}}

\put(20,1){\makebox(0,0)[t]{$\omega_r$}}
\put(1,15){\makebox(0,0)[r]{$\omega_i$}}

\end{picture}
\end{center}
\caption{Contour $\Gamma_2$}
\end{figure}
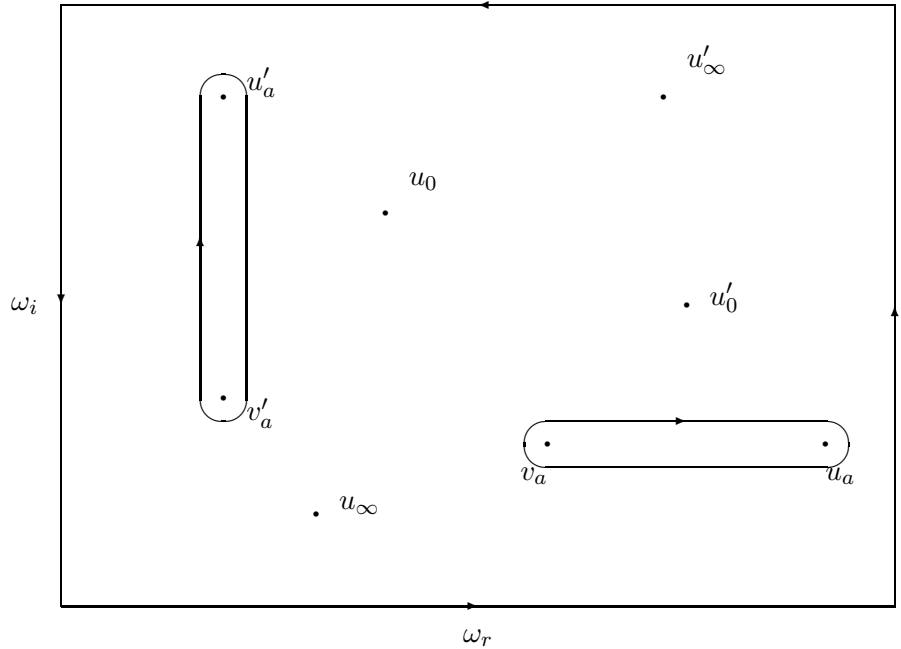
Using the transformation properties Eq.~(\ref{trans}) the integral along the boundary 
of the rectangle is easily computed to be 
\begin{equation}
-4\pi i\int_{\omega_r}\frac{du}{\zeta^{n-2}(u)} ,
\end{equation}
where $\int_{\omega_r}$ denotes the integral along the real period of $S$.
Integrals along the contours surrounding the pairs $(u_a, v_a)$ or $(u'_a, v'_a)$
get contributions only from the 
$\log\left(\sigma(u-u_a)\sigma(u-u'_a)/(\sigma(u-v_a)\sigma(u-v'_a))\right)$ 
part of $\log\frac{\rho}{\xi}$.
If one replaces the latter expression with 
$$\log\left(\sigma(u-u_a)\sigma(u-u'_a)\sigma(u-v_a)\sigma(u-v'_a)\right)$$ and simultaneously 
changes the contours to the figure-eight shaped contours in Figure~4, the value of
the integral is not changed. 
\begin{figure}[t]
\setlength{\unitlength}{0.8em}
\begin{center}
\begin{picture}(40,30)
%\put(8.5,0.5){\makebox(0,0){Figure~4}}
\put(2,2){\line(1,0){36}}
\put(38,2){\line(0,1){26}}
\put(38,28){\line(-1,0){36}}
\put(2,28){\line(0,-1){26}}

\put(9,11){\circle*{.2}}
\put(10,11){\makebox(0,0)[tl]{$v'_a$}}
\put(9,24){\circle*{.2}}
\put(10,24){\makebox(0,0)[bl]{$u'_a$}}
\put(13,6){\circle*{.2}}
\put(14,6){\makebox(0,0)[bl]{$u_{\infty}$}}
\put(16,19){\circle*{.2}}
\put(17,20){\makebox(0,0)[bl]{$u_0$}}
\put(23,9){\circle*{.2}}
\put(23,8){\makebox(0,0)[tr]{$v_a$}}
\put(35,9){\circle*{.2}}
\put(35,8){\makebox(0,0)[tl]{$u_a$}}
\put(29,15){\circle*{.2}}
\put(30,16){\makebox(0,0)[tl]{$u'_0$}}
\put(28,24){\circle*{.2}}
\put(29,25){\makebox(0,0)[bl]{$u'_{\infty}$}}

\put(32.5,9){\oval(7,2)}\put(32.5,10){\vector(1,0){0}}
\put(32,8){\vector(-1,0){0}}
\put(25.5,9){\oval(7,2)}\put(25.5,8){\vector(1,0){0}}
\put(25,10){\vector(-1,0){0}}
\put(9,21.25){\oval(2,7.5)}\put(8,21.25){\vector(0,1){0}}
\put(10,20.75){\vector(0,-1){0}}
\put(9,13.75){\oval(2,7.5)}\put(10,13.75){\vector(0,1){0}}
\put(8,13.25){\vector(0,-1){0}}

\put(20,2){\vector(1,0){0}}\put(38,15){\vector(0,1){0}}
\put(20,28){\vector(-1,0){0}}\put(2,15){\vector(0,-1){0}}

\put(20,1){\makebox(0,0)[t]{$\omega_r$}}
\put(1,15){\makebox(0,0)[r]{$\omega_i$}}

\end{picture}
\end{center}
\caption{Contour $\Gamma_a$}
\end{figure}
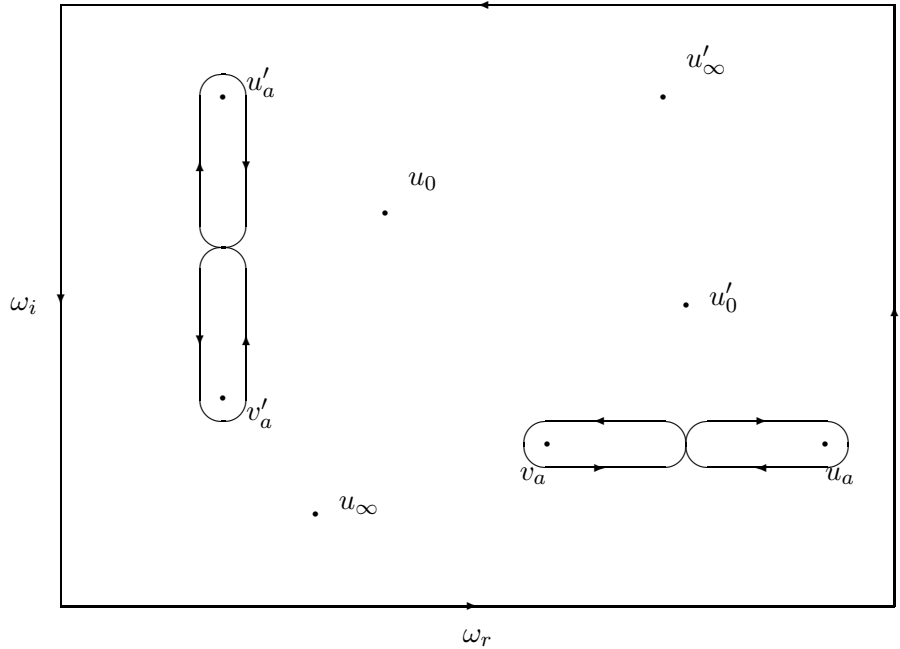
Finally the integral over the figure-eight shaped contour $\Gamma_a$ in Figure~4 is equal to
\begin{equation}
\oint_{\Gamma_a} \frac{du}{\zeta^{n-2}(u)} \log\left(\eta(u)-z_a(\zeta)\right).
\end{equation}
We must also evaluate the last term in Eq.~(\ref{zerostep}). This is easily done
using the definition of $\chi,\tilde{\chi}$ and the transformation rules 
Eqs.~(\ref{rhotrans},\ref{xitrans}).
Putting all this together we find
\begin{eqnarray}
\oint_{\! C}\frac{d\zeta}{\zeta^n}\hat{f}(\eta,\zeta)&=&4h\oint_0\frac{d\zeta}{\zeta^{n-1}}
-2\pi i\oint_{\omega_r}\frac{d\zeta}{\eta}
\zeta^{-n+2}\nn\\
&&+\frac{1}{2}\sum_a\oint_{\! C_a} \frac{d\zeta}{\eta}\zeta^{-n+2} \log(\eta-z_a(\zeta)).
\end{eqnarray}
Here $\eta=\sqrt{-\eta_2}$, and the contour $C_a$ in the $\zeta$ plane comes from the contour
$\Gamma_a$ in the $u$ plane. This implies
\begin{eqnarray} \label{Ffinal}\nonumber
F&=&\frac{1}{2\pi i}\oint_0 d\zeta \frac{4 h \eta_2}{\zeta^3}+2\oint_{\omega_r}d\zeta
\frac{\sqrt{-\eta_2}}{\zeta^2}-\\
&&\sum_a\frac{1}{2\pi i}\oint_{\! C_a}\ \frac{d\zeta}{\zeta^2}
(\sqrt{-\eta_2}-z_a(\zeta))\log (\sqrt{-\eta_2}-z_a(\zeta)).
\end{eqnarray}
The Legendre transform of this function with respect to $v,\bar{v}$
is the K\"ahler potential for the metric on $\M$.
Eq.~(\ref{Ffinal}) agrees with the conjecture of Chalmers in Ref.~\cite{Chalmers}. To see that
the metric is indeed ALF, one must consider the behaviour of $F$ at infinity, which
corresponds to taking $k_1$ to infinity. It is rather easy to see that in this limit the 
quartic polynomial $\eta_2(\zeta)$ tends to $-(P(\zeta))^2$, where $P(\zeta)$ is
a quadratic polynomial in $\zeta$ (this is a consequence of Eq.~(\ref{constrain})).
Substituting this into the formula for $F$ one can see that $F$ takes the Taub-NUT 
form~\cite{IR}, with $h$ being inversely proportional to the radius of the Taub-NUT
circle at infinity. Thus the metric is indeed ALF.

\section{Conclusions and open problems}
String-theoretic arguments indicate that gravitational instantons can be identified
as the moduli spaces of Nahm equations.
We used this relationship to construct $D_k$ ALF gravitational instantons and find their twistor 
spaces and K\"ahler potentials. Finding the K\"ahler potential explicitly requires solving a 
transcendental constraint $\partial F/\partial w=0$ and performing Legendre transform.
In the case of the Atiyah-Hitchin manifold ($D_0$ ALF space in our notation) this constraint can 
be solved~\cite{IR}. Solving the constraint in the general case seems to be hard.

In fact for $k\leq 4$ a description of $D_k$ ALF spaces as finite
hyperk\"ahler quotients of known hyperk\"ahler manifolds (Eguchi-Hanson manifold, 
Dancer's manifold~\cite{Dancer0} and ${\bf R}^4$) was
presented in Ref.~\cite{us1}. It would be interesting to compare the two descriptions and to 
see how in these cases the constraint $\partial F/\partial w=0$ is effectively solved.

In the limit $h\to 0$ our $D_k$ ALF metrics become ALE. On the other hand, 
Kronheimer constructed the same metrics as hyperk\"ahler quotients~\cite{ALE}.
Apparently, in Kronheimer's description the difficulty of solving the transcendental constraint
translates into the difficulty of solving a system of algebraic equations.

Hyperk\"ahler manifolds which have two compact directions at infinity are also of physical 
interest. It remains to be seen whether the string-theoretic approach can lead to a
construction of such manifolds.

\renewcommand{\thesection}{Acknowledgments}
\section{}

We would like to thank John H. Schwarz for reading the manuscript. 
S. Ch. is grateful to IAS for hospitality.

\end{document}